\documentclass[a4paper]{jpconf}
\usepackage{graphicx}

\begin{document}

\title{Electronic structure of cubic ScF$_3$ from first-principles calculations}


\author{P Zhgun$^1$, D Bocharov$^{1,2,3}$, S Piskunov$^{1,2,3}$, A Kuzmin$^3$ and J Purans$^3$}

\address{$^1$ Faculty of Physics and Mathematics, University of Latvia, Zellu Str. 8, LV-1002 Riga, Latvia}
\address{$^2$ Faculty of Computing, University of Latvia, Raina Blvd. 19, LV-1586 Riga, Latvia}
\address{$^3$ Institute of Solid State Physics, University of Latvia, Kengaraga Str. 8, LV-1063 Riga, Latvia}

\ead{pjotr.zgun@gmail.com}

\begin{abstract}
The first-principles calculations have been performed to investigate the ground state properties of cubic scandium trifluoride (ScF$_3$) perovskite. Using modified hybrid exchange-correlation functionals within the density functional theory (DFT) we have comprehensively compared the electronic properties of ScF$_3$ obtained by means of the linear combination of atomic orbitals (LCAO) and projector augmented-waves (PAW) methods. Both methods allowed us to reproduce the lattice constant experimentally observed in cubic ScF$_3$ at low temperatures and predict its electronic structure in good agreement with known experimental valence-band photoelectron and F 1$s$ X-ray absorption spectra.
\end{abstract}

\section{Introduction}
Scandium trifluoride (ScF$_3$) is a promising material attracted much attention due to recently discovered strong negative thermal expansion (NTE) coefficient over a wide range of temperatures from 10 to 1100~K \cite{JACS}.

\begin{figure}[t]
\includegraphics[width=18pc]{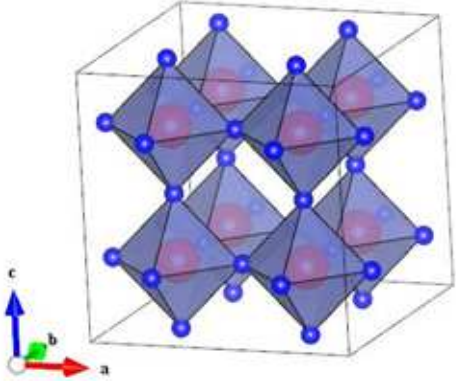}\hspace{2pc}%
\begin{minipage}[b]{17pc}\caption{\label{figure: structure}
 Schematic view of cubic  ScF$_3$ structure (space group $Pm\overline{3}m$) built up of ScF$_6$ regular octahedra joined by corners (small balls are fluorine atoms, large balls are scandium atoms). All Sc--F--Sc angles are equal to 180$^\circ$.}
\end{minipage}
\end{figure}

At atmospheric pressure ScF$_3$ has cubic  (space group $Pm\overline{3}m$) ReO$_3$-type structure  (Fig.~\ref{figure: structure}) down to at least 10~K \cite{JACS}.
However, opposite to the metallic ReO$_3$, which has very week NTE effect \cite{Chatterji2008,Rodriguez2009}, scandium trifluoride is an insulator with band gap of more than 8~eV \cite{Umeda}.
A cubic-to-rhombohedral phase transition occurs in ScF$_3$ at high pressure  ($P$$>$0.5~GPa at $T$$\sim$300~K or $P$=0.1--0.2~GPa at 50~K) as determined from X-ray and neutron diffraction studies \cite{JACS,Aleksandrov2011}. Raman spectroscopy confirms that upon pressure increase cubic ScF$_3$ undergo a phase transition to the rhombohedral (space group $R\bar{3}c$) phase  \cite{Aleksandrov2011,Aleksandrov2002}.

The electronic structure of ScF$_3$ thin films has been investigated by means of resonant photoemission spectroscopy at the Sc 2$p$  and F 1$s$ absorption edges in Ref.~\cite{Umeda}. Two charge-transfer-type satellites, the first one at about 13~eV below the main peaks in the valence-band and the second one with an energy separation of about 9--10~eV, were observed and indicate the strong hybridization effect between Sc 3$d$ and F 2$p$ states  \cite{Umeda}.

The phonon properties of ScF$_3$ have been studied using a combination of inelastic neutron scattering experiment with \textit{ab initio} calculations of lattice dynamics \cite{neutrons}. It was shown that a description of NTE within the quasi-harmonic approximation  is not reliable in the case of ScF$_3$. The authors in  \cite{neutrons} have  demonstrated that the $R4+$ mode (the one with the lowest energy at $R$-point of Brillouin zone (BZ)) has quartic potential and  proposed the mechanism for NTE based on this anharmonicity. The transition between the ground state and the first excited state ($\sim$19~meV) of the $R4+$ mode has been associated with experimentally observed phonon peak ($\sim$25~meV), which stiffens upon increasing temperature \cite{neutrons}.

In spite of mainly experimental efforts have been performed to understand the NTE in ScF$_3$, its origin is still under debate. Therefore, comprehensive theoretical study is necessary to elucidate the nature of this peculiar property. In this study we report the results of \textit{ab initio} electronic structure calculations for cubic ScF$_3$ using the linear combination of atomic orbitals (LCAO) and projector augmented-waves (PAW) methods.


\section{Computational details}
In this study ScF$_3$ is modeled by means of two different methods: (i) LCAO within the framework of hybrid density functional approach and (ii) PAW calculations using the generalized gradient approximation (GGA) density functional.

To perform hybrid LCAO calculations, we used the periodic CRYSTAL09 code \cite{CRYSTAL}, which employs Gaussian-type functions centered on atomic nuclei as the basis set for expansion of the crystalline orbitals. The full-electron basis sets used in this study for F and Sc were taken from CRYSTAL basis set library \cite{CRYSTAL}. Threshold parameters for evaluation of different types of bielectronic integrals such as overlap and penetration tolerances for Coulomb integrals, overlap tolerance for exchange integrals and pseudo-overlap tolerances for exchange integral series \cite{CRYSTAL} have been set to 10$^{-8}$, 10$^{-8}$, 10$^{-8}$, 10$^{-8}$ and 10$^{-16}$, respectively.
Calculations are considered as converged when the total energy obtained in the self-consistent field procedure differs by less than 10$^{-10}$~a.u. in the two successive cycles.

LCAO calculations of ScF$_3$ have been performed using a set of hybrid Hartree-Fock/Kohn-Sham (HF/KS) exchange-correlation functionals \cite{PBE,PBESOL,LYP,BECKE,PWGGA} combining exact HF nonlocal exchange and KS exchange operator within the GGA as implemented in the CRYSTAL09 code \cite{CRYSTAL}. We have constructed the following hybrid functionals: ``BECKE \& LYP'', ``BECKE \& PWGGA'', ``PBE \& PBE'', ``PBEsol \& PBEsol'', where the first terms refer to the KS exchange and the second terms to the KS correlation parts of a hybrid functional, while the weight of the HF exchange has been varied for each functional separately between 0 and 100\%.

 DFT calculations using the PAW method as implemented in the VASP code \cite{VASP} were performed for comparison. We have compared PW91 \cite{PWGGA} and PBE \cite{PBE} exchange-correlation potentials as well as large-core effective core potential (LC-ECP) with three 3$d^1$4$s^2$ valence electrons and small-core effective core potential (SC-ECP) with eleven 3$s^2$3$p^6$3$d^1$4$s^2$ valence electrons for Sc atom. Fluorine atom for all calculations was described as ECP with 2$s^2$2$p^5$ valence shell. Hybrid HF-DFT calculations have been also performed with the weigth of exact HF exchange part equal to 25\%. The optimal cut-off energy has been equal to 1000~eV.

To provide the balanced summation over the direct and reciprocal lattices of ScF$_3$ in both LCAO and PAW methods, the reciprocal space integration has been performed by sampling the BZ with the 8$\times$8$\times$8 Pack-Monkhorst $k$-mesh \cite{Monkhorst} that results in 35 eventually distributed $k$-points in the irreducible BZ.

\section{Choice of exchange-correlation functional}

\begin{figure}[t]
\includegraphics[width=20pc]{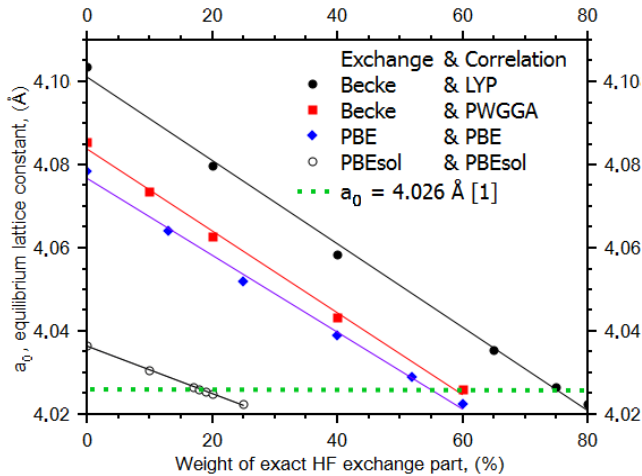}\hspace{2pc}%
\begin{minipage}[b]{15pc}\caption{\label{figure: latt_const}Equilibrium lattice constant with respect to the weight of exact HF non-local exchange in hybrid density calculation by means of the LCAO method.}
\end{minipage}
\end{figure}

The influence of the exchange-correlation functional type and the weight (\textit{w}HF) of exact HF non-local exchange in hybrid density calculations within the LCAO method on the lattice parameter in ScF$_3$ is shown in Fig.~\ref{figure: latt_const}.
We found that pure GGA functionals overestimate the experimental value of lattice constant $a_0$=4.026~\AA\ \cite{JACS}, while stepwise increase of the \textit{w}HF for about 10\% leads to proportional decrease of $a_0$ by approximately $-$0.01~\AA. The PBEsol functional with \textit{w}HF=18\% reproduces perfectly the experimental lattice parameter and has been selected for further calculations.

Values of the lattice constant $a_0$ obtained using pure GGA calculations by the VASP code \cite{VASP} are in the range of 4.02--4.03~\AA\ if LC-ECP is used for Sc, while $a_0$ of 4.06--4.07~\AA\ has been obtained if SC-ECP has been adopted for Sc. Using hybrid HF-DFT calculations with the PBE0 functional \cite{VASP}), the value of $a_0$
is equal to $\sim$4.04--4.05~\AA: it is larger when LC-ECP is used for Sc and is smaller when SC-ECP is used. 


To additionally justify the choice of hybrid functional, we have calculated the value of the band gap ($E_g$). Until now the band gap of ScF$_{3}$ was not precisely measured experimentally, but it is estimated to be larger than 7--8~eV \cite{Umeda}. All hybrid functionals constructed for LCAO method ($w$HF corresponds to $a_0$=4.026~\AA, see Fig.~\ref{figure: latt_const}) show that ScF$_{3}$ is an insulator with relatively large band gap \textit{E}$_{g}$ $>$ 8~eV. 
In the PAW GGA calculations the band gap is  underestimated  as it is predicted by the theory for pure GGA functionals:
it is equal to 5.5--6.0~eV. Hybrid PAW calculations allowed us to obtain $E_g$ equal to 8--9~eV, which  is closer to the experimental estimate.

Thus, for electronic structure calculations we have finally chosen the ``PBEsol \& PBEsol'' functional with $w$HF=18\%, giving $E_g$=9.8~eV, for LCAO method and the PW91 functional with $w$HF=25\%, giving $E_g$=8.9~eV, for PAW method.

\section{Electronic properties}
The effective charges of scandium and fluorine atoms were calculated using Mulliken population analysis \cite{CRYSTAL} in LCAO method and  Bader topological analysis \cite{Bader} in PAW calculations. The LCAO charges are equal to $q_{\rm Sc}$=$+$2.28e and $q_{\rm F}$=$-$0.76e.
In PAW method the atom charges were slightly different for two Sc pseudopotentials: $q_{\rm Sc}$=$+$2.70e and $q_{\rm F}$=$-$0.90e for LC-ECP and
$q_{\rm Sc}$=$+$2.17e and $q_{\rm F}$=$-$0.73e for SC-ECP.
 These results demonstrate a considerable deviation of effective atomic charges from formal charges caused by partly covalent nature of Sc--F bonds.


\begin{figure}[t]
\includegraphics[width=22pc]{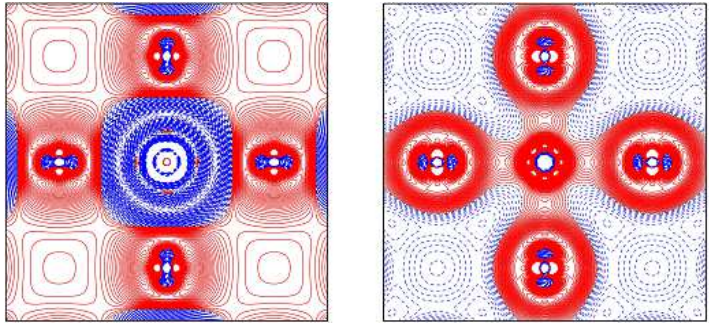}\hspace{2pc}%
\begin{minipage}[b]{13pc}\caption{\label{figure: ECHD}Plots of the charge density difference in (100) plane between the full charge density of crystal and electronic density of non-interacting atoms (left panel) and ions (right panel) obtained using ``PBEsol \& PBEsol'' functional with $w$HF = 18\%. Solid (red) and dashed (blue) isolines correspond to positive (excess) and negative (deficiency) electron density difference, respectively.}
\end{minipage}
\end{figure}

\begin{figure}[t]
\includegraphics[width=22pc]{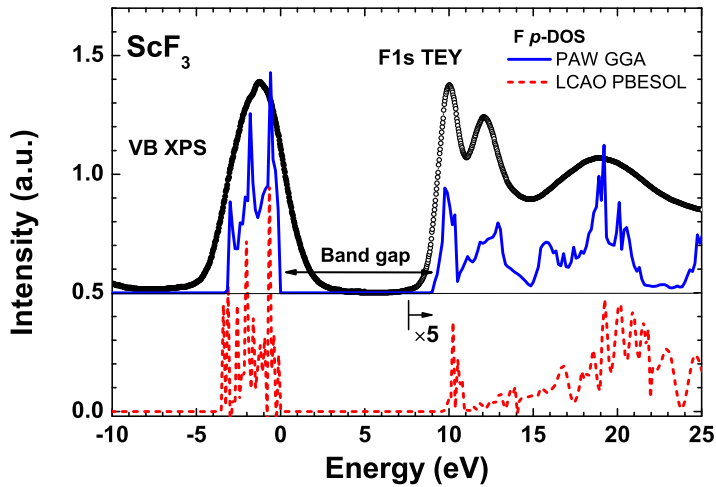}\hspace{2pc}%
\begin{minipage}[b]{13pc}\caption{\label{figure: fig5} A comparison of the F $p$ density of states (DOS) calculated by means of the LCAO (dashed line) and PAW (solid line)  methods with valence-band photoelectron \protect\cite{SSC1993} and F 1$s$ X-ray absorption spectra \protect\cite{Umeda}. The amplitude of the DOS in the conduction band was multiplied by a factor of 5.
Energy scale is given with respect to the calculated top of the valence band.
Experimental spectra has been aligned to match best the calculations.}
\end{minipage}
\end{figure}

Fig.~\ref{figure: ECHD} shows electron density maps in (100) plane of ScF$_3$ obtained using LCAO calculation as a difference between the full charge density of crystal and electronic density of non-interacting atoms or ions. The anisotropy of electron density induced by bond covalent contribution is clearly visible in both graphs: the electron density is larger along the crystallographic axes.

In Fig.~\ref{figure: fig5} we compare the partial F $p$ density of states (DOS) calculated by both LCAO and PAW methods with the experimental valence-band (VB) photoelectron spectrum taken from Ref.~\cite{SSC1993} and the F 1$s$ X-ray absorption spectrum taken from Ref.~\cite{Umeda}. The VB photoelectron spectrum of ScF$_3$ has a complex shape composed of the main peak and the shoulder located at the left side, which originate due to the valence band splitting upon hybridization of the F 2$p$ and Sc 3$d$ states \cite{SSC1993}. Our theoretical calculations confirm such interpretation and reproduce well the width and asymmetric shape of the valence band. The F 1$s$ absorption spectrum \cite{Umeda} consists of three well visible peaks at 9.3, 11.4 and 18.2~eV due to transitions from the F 1$s$ core state to the F n$p$ states in the conduction band and above. The first two peaks correspond to the F 2$p$ states in the conduction band, which are hybridized with the Sc 3$d$ states. The two peaks originate from the $t_{2g}$-$e_g$ splitting of the Sc 3$d$ states by the octahedral crystal field. The third peak at 18.2~eV corresponds to the interaction of the F 2$p$ and Sc 4$p$ states. Such interpretation is fully supported by the calculated F $p$ DOS: in fact, good agreement with the F 1$s$ absorption spectrum  indicates that a relaxation
of the electron states due to the F 1$s$ core-hole is weak in ScF$_3$.

\section{Conclusions}
The lattice constant, band gap and electronic properties of cubic ScF$_{3}$  were calculated using LCAO and PAW DFT computational methods. It was shown that a variation of exact HF exchange part allowed
us to reproduce experimental value of the lattice constant $a_0$=4.026~\AA\ \cite{JACS}.

The band gap obtained in hybrid HF-DFT calculations is equal to 8--10~eV and agree well with the expected value \cite{Umeda}. A deviation of atom charges from formal ones and an anisotropy of electronic density  point to the covalent nature of Sc--F bonds in ScF$_3$. Calculated partial F $p$ DOS is in a good agreement with valence-band photoelectron  \cite{SSC1993} and F 1$s$ X-ray absorption \cite{Umeda} spectra.

Good agreement between the theory and available experimental data makes promising the application of the present
approach to the investigation of phonon properties in ScF$_3$ in the future work.

\ack
Authors are greatly indebted to R.A. Evarestov, D. Gryaznov, V. Kashcheyevs and A.I. Popov for many stimulating discussions. This work was supported by ESF Projects 2009/0202/1DP/1.1.1.2.0/09/APIA/VIAA/141 and 2009/0216/1DP/1.1.1.2.0/09/APIA/VIAA /044 and scholarship ESS2011/78/MS/6 within 2009/0162/1DP/1.1.2.1.1/09/IPIA/VIAA/004.

\end{document}